\documentclass[11pt,a4paper]{article}
\usepackage[hyperref]{emnlp2018}
\usepackage{times}
\usepackage{latexsym}

\usepackage{url}

\usepackage{url}

\usepackage{graphicx}
\usepackage{epstopdf}
\usepackage[vlined,ruled]{algorithm2e}
\usepackage{multirow}
\usepackage{amsmath}

\aclfinalcopy 

\SetKwProg{Fn}{Function}{}{}
\SetKwFor{For}{for (}{) $\lbrace$}{$\rbrace$}
\SetKwInOut{Parameter}{parameter}

\title{Hypernyms Through Intra-Article Organization in Wikipedia}

\author{Disha Shrivastava \thanks{* Work done as part of IBM Research, Bangalore} \\
  MILA, Universit\'e de Montr\'eal\\
   Montreal, Canada \\
  {\tt dishu.905@gmail.com} \\\And
  Sreyash Kenkre \\
  IBM Research\\
   Bangalore, India \\
  {\tt srekenkr@in.ibm.com} \\\And
  Santosh Penubothula \\
  IBM Research\\
   Bangalore, India \\
  {\tt sapenubo@in.ibm.com} \\}

\date{}

\begin{document}
\maketitle
\begin{abstract}
We introduce a new measure for unsupervised hypernym detection and directionality.
The motivation is to keep the measure computationally light and portatable across languages. We show that the relative physical location of words in explanatory articles captures the directionality property. Further, the phrases in section titles of articles about the word, capture the semantic similarity needed for hypernym detection task. We experimentally show that the combination of features coming from these two simple measures suffices to produce results comparable with the best unsupervised measures in terms of the average precision.
\end{abstract}

\section{Introduction}
Given two words $w_1$ and $w_2$, the hypernym detection task is to determine if there is a hypernym relation between the two words. If a hypernym is known to exist, the directionality task is to determine if $w_1$ is a hypernym or hyponym of $w_2$. More precisely, due to polysemy, the detection task asks if, there is some meaning of $w_1$, in which it is a hypernym or hyponym of some meaning of $w_2$.

The first approaches were pattern based \cite{Hearst:1992, Snow:2004}. However, these suffered from poor recall. This led to the development of methods based on the {\it distributional hypothesis} \cite{Harris:1954} or the {\it  Distributional Inclusion Hypotheses} \cite{GeffetD:2005}. The method used in these techniques was to take a very large corpus, and using either window based, or dependency path based approaches, along with measures like frequency, PPMI \cite{Church:1990}, LPMI \cite{Evert:2005}, to find vectors to represent the words. In supervised settings, the vectors for two words are combined suitably and a classifier is trained \cite{BaroniBDS:2012, RollerEB:2014, WeedsCRWK:2014, ShwartzGD:2016} to predict the existence of a hypernym relation and later directionality. However, recently there has been deeper research on what exactly is learned by these techniques \cite{LevyRBD:2015}. In the unsupervised setting a suitable measure, motivated by either the distributional inclusion hypothesis or the distributional informativeness hypothesis is used for hypernyms \cite{SantusLCLH:2016, WeedsWM:2004, SantusLLS:2014, GeffetD:2005}. 
\vspace{-0.45mm}

In this paper we present a simple and computationally light unsupervised technique for hypernym detection and directionality which is a combination of two measures, called as {\it depth measure} and {\it heading measure}. We start off with a large corpus, but instead of finding window or dependency path based contexts, which are very expensive to compute; we argue that the {\it internal organization of descriptive and explanatory documents} naturally leads to strong signals that are indicative of hypernyms. By exploratory documents, we mean documents that have been produced with the express purpose of making the reader understand the concepts that the document is describing; text books, research papers and Wikipedia articles are prime examples of this. We verify this intuition empirically, wherein we achieve results comparable and in some cases better than prior techniques in both the tasks of hypernym detection and directionality. One salient feature of our measures is that they exploit how humans organize information in explanatory documents making them portable across all languages. This offers us an advantage over prior techniques which depend on the intricacies of syntax and semantics of the language of the documents. 
 
\section{Methodology}
We will be using Wikipedia as the source of descriptive and explanatory documents in this paper. For ease of exposition, we define the concept of {\it units}. Given an article from Wikipedia, each page title, section title, sub-section title etc., irrespective of the depth of section, will be referred to as a {\it heading}. Each heading on the article usually consists of a title that describes what the text following the heading is about. Following the heading, are usually a few paragraphs that describe in more detail the heading. This may be followed by another heading, and this pattern repeats. We refer to each heading and the text following it, till (but not including) the next heading, as a unit. Thus the article is physically organized as a sequence of disjoint units. We represent a unit $u$ as a pair $(h, S)$, where $h$ is the heading, and $S$ is the sequence of sentences in the unit. 

\subsection{Depth Measure}
Given a hypernym-hyponym pair, $(w_1, w_2)$, 
consider the organization of a Wikipedia article containing both of them.
The very first unit at the top of the page is usually a broader introduction of the main topic of the article. It will use words that are more popular. However, from the next unit onward, the articles tend to be more specialized with higher detail in content. Thus the linguistic contexts used in the units occurring lower in the article tend to be more detail oriented than those occurring earlier (except for the first introductory unit). Since more detailed context are indicative of a hyponym, they tend to occur later in the article than the hypernyms. This same reasoning applies within a unit. In this case the hypernym will tend to occur in earlier sentences in the unit than the hyponym. We generalize this to the case in which $w_1$ and $w_2$ do not co-occur in the same article, as follows. We take a large corpus of articles (e.g. all articles in Wikipedia), and check the depth at which $w_1$ and $w_2$ tend to occur (individually). If $w_2$ tends to occur at larger depth than $w_1$, we conclude that $w_2$ is a hyponym of $w_1$. 

Let $\mathcal{P}$ be the set of articles. Let $a \in \mathcal{P}$ be an article, and let $w$ be a given word or phrase. To formally define depth, we will assume that the article has a fixed rooted tree like topology with the units of $a$ as its vertices, denoted by $G(a)$. The root will be the first unit of the article, and the depth will be the distance from the root. We experiment when $G(a)$ is a Star-like tree topology, as indicated by the depths of its sections and sub-sections, or a Linear-like topology with a unit being a parent of the immediate next unit in the physical layout of the article. We define a function $\lambda(a,w)$ that captures the depth of each occurrence of $w$ in $a$. 
Let $\mathcal{I}(a,w)$ denote the set of occurrences of $w$ in $a$.
Each occurrence consists of a pair $(u_i, s_j)$, where $u_i$ denotes a unit, and $s_j$ is the sentence in which it occurs. Multiple instances of $w$ in the same sentence is treated as one instance. Let $d(G(a))$ denote the total depth of $G(a)$.
If $d(u_i)$ is the depth of unit $u_i$ in $G(a)$, and $|u_i|$ is the number of sentences in it, then we define the
\begin{eqnarray}
& \lambda(a,w) = \sum\limits_{(u_i,s_j) \in \mathcal{I}(a,w)} \left(1-\frac{d(u_i)}{d(G(a)}\right)\left(1-\frac{j}{|u_i|}\right)& \nonumber
\end{eqnarray}

The first factor gives a normalized measure (to ensure same scale across all articles, of different sizes) of the depth of each occurrence of $w$ in $a$. Similarly, the second factor gives a normalized depth of the instance within a unit. Larger the $\lambda(a,w)$, more likely is it to be a hypernym. To aggregate this measure across all articles:

\begin{equation}
\lambda(w) =  \underset{a \in \mathcal{P}}{\mathrm{median}}\hbox{~} \lambda(a,w)
\end{equation}

\subsection{Heading Measure}
For testing relatedness between words, we define the heading measure, inspired by ~\cite{DoR:2012}. We search in Wikipedia for the article on the given phrase $w$ (e.g., if $w$ is the word {\it jumping}, then we get the article {\small \tt https://en.wikipedia.org/wiki/Jumping}). Since the page is about $w$, it is organized into sections that explain every property of $w$. We can thus represent $w$ simply by the collection of headings (titles, sub-titles at every possible level). 

If the page on $w$ turns out to be a disambiguation page, then the page lists different possible meanings of $w$, along with the corresponding links. We follow each of the links to get possible articles on different meanings of $w$. In case any of the pages is again a disambiguation page, we iterate further. For each of the pages that are articles, we form a set of headings. Each set corresponds to a different meaning of $w$. We let $S_w$ denote the collection of the headings for different meanings of $w$ (See Algorithm \ref{alg:ExtractHeadings}). Note here, that $S_w$ is a {\it set of sets}. One advantage of this method is that we get the different meanings of the words up front, whereas, in context feature based approaches, there can be a mixing of the different contexts for polysemous words.

\begin{algorithm}
\label{alg:ExtractHeadings}
\DontPrintSemicolon
\SetAlgoLined
\SetKwInOut{Input}{input}
\SetKwInOut{Output}{output}
\Input{Word or phrase $w$}
\Output{$S_w$, a collection of headings of pages on $w$}
\Fn{ExtractHeadings(w)}{
$\mathcal{P} = \{P_1,\ldots,P_k\}$ be the set of articles on $w$, $S_w \leftarrow \phi$  \;
\While{$\mathcal{P} \not=\phi$}{
   Select any $P \in \mathcal{P}$\;
   $\mathcal{P} \leftarrow \mathcal{P}\setminus P$\;
   \eIf{$P$ is not a disambiguation page}{
       Let $C$ be the collection of headings on page $P$\;
       $S_w \leftarrow S_w \cup \{ C \}$\;
      }{
       Let $\mathcal{D}$ be the collection of articles that $P$ points to as possible meanings of $w$\;
       $\mathcal{P} \leftarrow \mathcal{P} \cup \mathcal{D}$
       }
 }
 \Return $S_w$\;
 }
 \BlankLine
\caption{Extract Heading Sets}
\end{algorithm}

After computing $S_{w_1}$ and $S_{w_2}$ as shown in Algorithm ({\ref{alg:ExtractHeadings}), we compute the $SimScore(w_1, w_2)$ as the maximum similarity between an element of $S_{w_1}$ and $S_{w_2}$. For the similarity, we experimented with two measures, the {\it Jaccard Similarity}, and the {\it cosine} of the corresponding {\tt word2vec}~\cite{MikolvoCCD:2013} vectors. For using {\tt word2vec}, for each heading set $C$, we take the mean of the vectors for each heading. Since {\tt word2vec} uses the context of words, this combines our features with the contextual features. The final measure we use for the pair of words is:

\begin{eqnarray}\label{fullScore}
& \left( \frac{1 + \lambda(w_1)-\lambda(w_2)}{2}\right)\left(SimScore(w_1, w_2)  \right) &
\end{eqnarray}

\section{Experiments and Results}
\subsection{Datasets and Corpus}
We experimented with four datasets widely used in literature: BLESS~\cite{BaroniL:2011}, EVALution~\cite{SantusYLH:2015}, Lenci/Benotto~\cite{Bentto:2015}, and Weeds~\cite{WeedsCRWK:2014} taken from the repository provided by \cite{SantusSS:2017}. 
The corpus of articles we use is a complete {\tt xml} dump of the English Wikipedia dated {\tt 3 Nov 2017}. 

\subsection{Testing Directionality}
We extracted out the pairs marked hypernyms from each of the four data sets and computed the depth measure for each word in the pair. If the difference $\lambda(w_1)$ - $\lambda(w_2)$ is less than zero, we mark these pairs as False and compute precision. 
To identify the articles containing $w$, we indexed the corpus of Wikipedia using Elasticsearch \cite{Gormley:2015:EDG:2904394} and used the top thousand articles returned as the set for computing $\lambda(w)$. We experimented with the Star and Linear topologies. 

\begin{table*}
\footnotesize
\centering
\footnotesize
\begin{tabular}{|c|c|c|c|}
\hline
{\bf Dataset}		 &	&{{\bf Star Topology}}&{{\bf Linear Topology}} \\\hline			
					&	Total&	Precision&  	Precision	 \\\hline
BLESS		 		&	1198 &	0.918	 &		0.536	 	\\\hline
Weeds		 		&	1321 &	0.974	 &		0.429	 	\\\hline
Evalution	 		&	3303 &	0.980	 &		0.566	 	\\\hline
LenciBenotto 		&	1728 &	0.974	 &		0.439	 	\\\hline
\end{tabular}
\caption{Testing directionality. Total= total number of pairs present in the test set. }
\label{table:DirectionalityResults}
\end{table*}

It is seen that with Star topology our precision in very high on each of the datasets. The worst performance is on BLESS. However, here too, it is $91.8\%$. For BLESS, as seen in ~\cite{SantusLLS:2014}, the performance of SLQS is $87\%$. 
Similar to SLQS, our depth measure is motivated by {\it distributional informativeness hypothesis}~\cite{SantusSS:2017}. However, without using the extensive computation of context vectors and entropy, we are able to demonstrate good performance. As can be seen by physically examining Wikipedia articles, many of them tend to have a Star topology. This is also indicative that the topology used plays a major role in this feature. More sophisticated techniques will be needed to identify the topology of individual articles.

\subsection{Testing Detection}

For this experiment, we aim to discriminate pairs of words connected by the hypernym relation, from words connected by other relations (meronym, coord, attribute, event, antonym, synonym). For each pair, we evaluate our scoring function given in expression (\ref{fullScore}). We compared our numbers with those given in ~\cite{SantusSS:2017}. In that paper, multiple measures are used, and the best performing measure for every row of the table is presented. We conducted the experiments for both, Star as well as the Linear topology. However, the results for Star topology were slightly better, hence we present these in Table (\ref{table:DetectionExptJaccard}).

\begin{table*}[t]
\footnotesize
\centering
\footnotesize
\begin{tabular}{|c|c|c|c|c|c|}
\hline
{\bf Dataset} 					& {\bf Hyper vs Rel} 	& {\bf AP word2vec} 	& {\bf AP Jaccard} 			& {\bf Best AP} 	& {\bf Best Measure}\\ \hline		
\multirow{5}{*}{BLESS}			& all other relations	& {\bf 0.084}						& {\bf 0.065}							& 0.051					&  $invCL$			\\ \cline{2-6}			
								& meronym				& 0.446						& 0.355							& 0.760					&  $SLQS_{sub}$		\\ \cline{2-6}				
								& coord					& 0.203						& 0.235							& 0.537					&  $SLQS_{sub}$		\\ \cline{2-6}				
								& attribute				& 0.581						& 0.509							& 0.740					&  $SLQS_{sub}$		\\ \cline{2-6}				
								& event					& 0.453						& 0.315							& 0.779					&  $APSyn$			\\ \hline		
\multirow{2}{*}{Weeds}			& all other relations	& {\bf 0.514}						& {\bf 0.506}	  						& 0.441					&  $clarkeDE$		\\ \cline{2-6}								
								& coord					& {\bf 0.514}						& {\bf 0.506}	  						& 0.441					&  $clarkeDE$		\\ \hline								
\multirow{5}{*}{EVALution}  	& all other relations	& 0.273						& 0.290	 						& 0.353					&  $invCL$			\\ \cline{2-6}								
								& meronym				& 0.629						& {\bf 0.678}	 						& 0.675					&  $APSyn$			\\ \cline{2-6}								
								& attribute				& 0.556						& 0.614	 						& 0.651					&  $APSyn$			\\ \cline{2-6}								
								& antonym				& 0.520						& 0.526	 						& 0.550					&  $SLQS-row$		\\ \cline{2-6}								
								& synonym				& 0.606						& 0.593	 						& 0.657					&  $SLQS-row$		\\ \hline								
\multirow{3}{*}{Lenci/Benotto}	& all other relations	& {\bf 0.401}						& {\bf 0.389}	 						& 0.382					&  $APSyn$			\\\cline{2-6}								
								& antonym				& 0.548						& 0.530	 						& 0.624					&  $APSyn$			\\\cline{2-6}							
								& synonym				& 0.599						& 0.593	 						& 0.725					&  $SLQS-row_{sub}$	\\ \hline								
		 
\end{tabular}
\caption{$AP=$ average precision. The Best AP and Best Measure is taken from ~\cite{SantusSS:2017}.}
\label{table:DetectionExptJaccard}
\end{table*}

For the case of hypernym vs all other relations, except for EVALution, in all other data sets, our average precision ($AP$) using both Jaccard and word2vec (\cite{SantusSS:2017} call this as $AP$@all) is better than the best unsupervised measure as reported in~\cite{SantusSS:2017}. For comparing hypernyms against individual relations, we find that with Jaccard similarity, it performs better than the best measures on meronyms in EVALution, and coordinates in Weeds. However, it performs worse for both the relations in BLESS. Our systems performs worse than the best measure whenever an {\it Informativeness Measure} ~\cite{SantusSS:2017}, like $SLQS$ and its variants perform well. It performs better, or at least competitive, when the best performing measure is an {\it Inclusion Measure} or {\it Similarity Measure} (except for hypernym-vs-event in BLESS). A possible explanation of this is that the heading features that we use do not capture how informative a phrase is. However, having common headings is an indication of shared features, implying similarity, which is also indicated by inclusion measures. However, it should be noted, that {\it we are comparing our single system against the best performing one in each case}. For finding the best measure,~\cite{SantusSS:2017} finds the best by varying the measures as well as the features, whereas we have a fixed system. Our system took a day to set up (including coding effort), and a few mins to run. This is in contrast to methods mentioned above that rely on the computation of context vectors; calculation of dependency parse tree based features alone, from ukWack and Wackypedia corpus, took several days on the same machine.

\subsection{Error Analysis}
One of the sources of error in our technique is a {\it semantic drift} due to disambiguation pages. For example, for the pair $(alligator, wild)$, which is
marked as attribute in BLESS, our system follows disambiguation links to {\it wildlife}, and then marks it as a hypernym. We find this pattern repeatedly. e.g. $(scale, lizard)$ is a meronym in BLESS, but is classified as hypernym in the hypernym vs. meronym experiments. While $(scale, snake)$ is a meronym in BLESS, and is marked correctly as {\it not} a hypernym. One reason for this is that among the disambiguation pages, a word is often generalized to related terms. For the hypernym v/s all experiment, the proportion of false positives when at least one word needed a disambiguation page was $37\%$ for BLESS, $29\%$ for Weeds, and about $31\%$ for EVALution and Lenci/Benotto. Selective link following during the disambiguation step can potentially solve this problem.
%

\section{Conclusion}
We showed that the organization of articles is an important feature for the task of both hypernym detection and directionality. Using just this simple and computationally cheap measure suffices to give performance that
is comparable to the state of art unsupervised measures in these tasks. The proposed measure can also be trivially extended to any languages with a Wikipedia. We believe future work in this area will benefit by using this feature in complex systems that can improve performance.

\bibliography{emnlp2018}
\bibliographystyle{acl_natbib_nourl}

\end{document}